\documentclass[aps, prd, superscriptaddress, twocolumn, a4paper]{revtex4}
\usepackage{graphicx}
\usepackage[sort&compress]{natbib}
\usepackage{epsfig}
\usepackage{amsmath}

\def\be{\begin{equation}}
\def\ee{\end{equation}}

\begin{document}

\title{Strangeness enhancement and flow-like effects in $e^+e^-$ annihilation at high parton density}

\author{P. Castorina}
\affiliation{INFN, Sezione di Catania, I-95123 Catania, Italy.}
\affiliation{Institute of Particle and Nuclear Physics, Faculty of Mathematics and Physics, 
Charles University, V Hole\v{s}ovi\v{c}k\'ach 2, 18000 Prague 8, Czech Republic.}

\author{D. Lanteri}
\affiliation{INFN, Sezione di Catania, I-95123 Catania, Italy.}
\affiliation{Dipartimento di Fisica e Astronomia, Università  di Catania, I-95123 Catania, Italy.}

\author{H. Satz}
\affiliation{Fakult\"at f\"ur Physik, Universit\"at Bielefeld, Germany.}

\date{\today}
\begin{abstract}
Strangeness enhancement and collective flow are considered signatures of the
quark gluon plasma formation. These phenomena have been detected not only in relativistic heavy ion collisions
but also in high energy, high multiplicity events of proton-proton  and proton-nucleus (``small systems'') scatterings. Indeed, a universal behavior emerges by considering the parton density in the transverse plane as the dynamical quantity to specify the initial condition of the collisions.
On the other hand,  $e^+e^-$ annihilation data at LEP and lower energies indicate that there is no strangeness enhancement and no flow-like effect. We show that the parton density in the transverse plane generated in  $e^+e^-$ annihilation at the available energy is too low to expect such effects. The
event-by-event multiplicity where strangeness suppression and flow-like phenomenon could show up in $e^+e^-$ is evaluated. 

\end{abstract}

\pacs{04.20.Cv,11.10.Wx,11.30.Qc}
\maketitle

\section*{Introduction}

Recent experimental results in
proton-proton ($pp$) and proton-nucleus ($pA$) collisions~\cite{ALICE:2017jyt,Abelev:2013haa,Khachatryan:2016txc,atlas1,cms1,cms2,PHENIX:2018lia,p1,p2} support the conclusion that the
system created in high energy, high multiplicity
collisions with ``small'' initial settings ($pp$ and $pA$), is essentially the same as that one produced
with ``large'' initial nucleus-nucleus ($AA$) configurations.

The ALICE collaboration reported~\cite{ALICE:2017jyt} an enhanced production
of multi-strange hadrons, previously observed in $PbPb$
collisions~\cite{ABELEV:2013zaa}, in high energy, high multiplicity $pp$ events. 

Moreover, the energy loss in $AA$ collisions was
shown to scale in small and large systems~\cite{loss} by considering a
dynamical variable,  previously introduced to predict the strangeness
enhancement in $pp$~\cite{cs1,cs2,cs3,cs4}. It  corresponds to the initial entropy 
density of the collisions and takes into account the
transverse size (and its fluctuations) of the initial
configuration in high multiplicity events.

Another important similarity among $pp$, $pA$, and $AA$
collisions was identified in several measurements of long-range di-hadron azimuthal correlations~\cite{Aad:2012gla,atlas1,cms2,Khachatryan:2016txc}  indicating universality in
flow-like patterns.

More recently,  the $e^+ e^-$
annihilation LEP data have been reconsidered~\cite{badea} to check if a flow-like behavior is generated with this initial, small, non hadronic, setting.

The answer is negative and confirmed at lower energy by the BELLE collaboration~\cite{belle}.

Furthermore, there is no strangeness enhancement in $e^+ e^-$ annihilation. Figs.~(\ref{Fig:2},~\ref{Fig:4}) show the strangeness suppression factor, $\gamma_s$, in the Statistical Hadronization model (SHM) 
\cite{Becattini:2009sc,pbm,gammasnoi} as a function of the available energy ($\gamma_s \simeq 1$ means no suppression, i.e. enhancement).

According to the universality point of view, strangeness enhancement and
collective flow are both indications of the formation of an initial system with high entropy density, i.e. high  parton number density in the transverse plane.

In this letter we show that at the available energies in $e^+e^-$ annihilation
the parton density in the transverse plane is small and therefore the previous
signatures should not arise here.

The event-by-event multiplicity and the corresponding energy where strangeness suppression and flow-like phenomenon could show up in $e^+e^-$ turns out to be quite large.

In the next section the universality in hadronic and nuclear collision will be recalled. Sec.~\ref{sec:2} is devoted to evaluate the parton density in the transverse plane for $e^+e^-$ annihilation and to verify that
the energy/multiplicity is actually too low to follow the universal trend, observed in the small and large hadronic and nuclear systems.  Sec.~\ref{sec:CC} contains comments and conclusions.

\section{\label{sec:1}Universality in hadronic and nuclear collisions}

\subsection{Strangeness enhancement}

\begin{figure}	
	\includegraphics[width=\columnwidth]{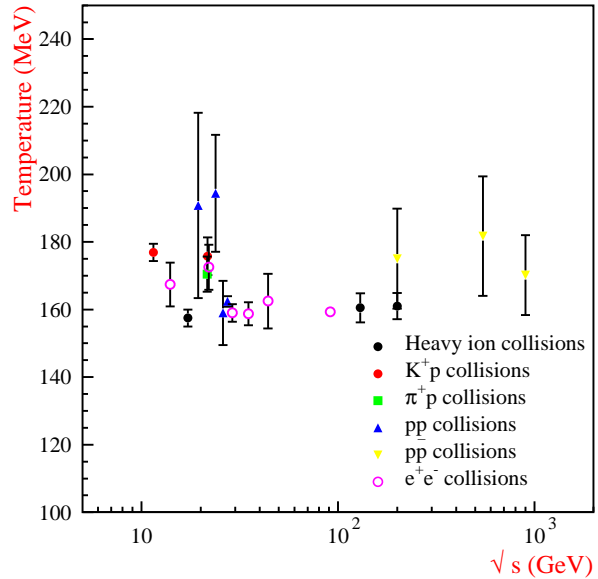}
	\caption{Hadronization temperature in the SHM} 
	\label{Fig:1}
\end{figure}

One of the most striking observations in high energy multihadron production is that both species
abundances and transverse momentum spectra (provided effects of collective flow and gluon radiation
are removed) follow the thermal pattern of an ideal hadron-resonance gas, with a universal temperature
$T \simeq 150 \pm 10$ MeV~\cite{pbm,Becattini:2009sc}(see fig.~\ref{Fig:1}).

More precisely, the relative yields of the different hadron species  are well
accounted for by an ideal gas of all hadrons and hadronic resonances
with one well-known caveat: strangeness production is
reduced with respect to the predicted rates. This suppression can, however, be taken
into account by one further parameter, $0< \gamma_s \le 1$.
The predicted rate for a hadron
species containing $\nu = 1$, $2$, $3$ strange quarks is then suppressed by the factor $\gamma_s^\nu$~\cite{Letessier:1993qa}. 

The basic quantity for the resonance gas description is the grand-canonical
partition function for an ideal gas at temperature $T$ in a spatial volume $V$
\be
\ln Z(T) = V \sum_i {d_i \gamma_s^{\nu_i}\over (2\pi)^3}~\! \phi(m_i,T),
\ee
with $d_i$ specifying the degeneracy (spin, isospin) of species $i$,
and $m_i$ its mass; the sum runs over all species. Here
\begin{equation}
\begin{split}
\phi(m_i,T) =& \int d^3p ~\exp\left\{\frac{\sqrt{p^2 + m_i^2}}{T}\right\}
\simeq\\
&\simeq \exp\left\{-\frac{m_i}{T}\right\}
\end{split}
\end{equation}
is the Boltzmann factor for species $i$, so that the ratio of the
production rates $N_i$ and $N_j$ for hadrons of species $i$ and $j$ is 
given by
\be
{N_i\over N_j}= {d_i \gamma_s^{\nu_i}\phi(m_i,T) 
\over d_j \gamma_s^{\nu_j}\phi(m_j,T)},
\ee
where $\nu_i=0$, $1$, $2$, $3$ specifies the number of strange quarks in species $i$.
We note that in the grand-canonical formulation the volume cancels out in
the form for the relative abundances.

The Statistical Hadronization model (SHM) is in agreement with the high energy data for large (nucleus-nucleus) and small (proton-proton and $e^+e^-$) initial settings with the same hadronization temperature as shown in fig.~\ref{Fig:1}~\cite{becattiniA,pbm}.

\begin{figure}	
	\includegraphics[width=\columnwidth]{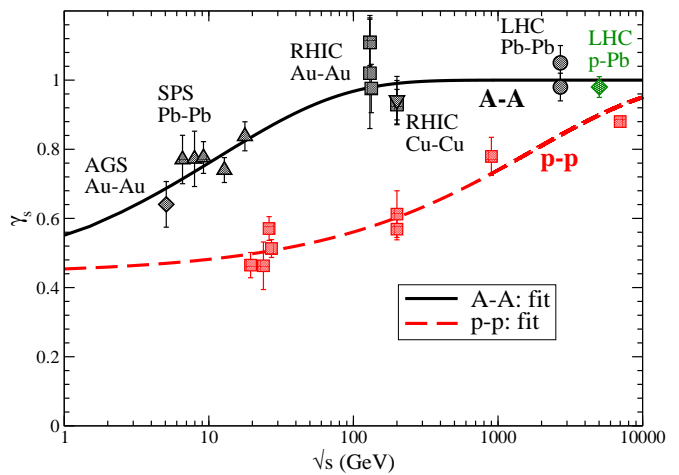}
	\caption{$\gamma_s$: energy dependence for $pp$, $pA$, $AA$ } 
	\label{Fig:2}
\end{figure}

It was  early proposed, on perturbative QCD basis, that the quark-gluon plasma
formation would enhance strange particle production in nucleus-nucleus $(AA)$ collisions~\cite{Rafelski:1982pu,Rafelski:1983hg}. Indeed,  $\gamma_s \simeq 1$ well describes the high energy $AA$ data, but
$\gamma_s < 1$ for proton-proton scattering at energies less than those at the Large Hadron Collider (LHC). Fig.~\ref{Fig:2} shows $\gamma_s$ as a function of the collision energy.

More recently, the ALICE collaboration reported  for $pp$ collisions~\cite{ALICE:2017jyt} the enhanced production of multi-strange hadrons,  previously observed in $PbPb$ collisions, in high energy, high multiplicity, proton-proton $pp$ events. Indeed a universal behavior of strangeness production, suggested on theoretical grounds in refs.~\cite{cs1,cs2,cs3,cs4,nuovo,floris}, emerges 
by considering a specific dynamical variable corresponding to the parton density in the transverse plane of the collision, which takes into account the transverse size (and its fluctuations)  of the initial configuration in high multiplicity events.

\begin{figure}	
	\includegraphics[width=\columnwidth]{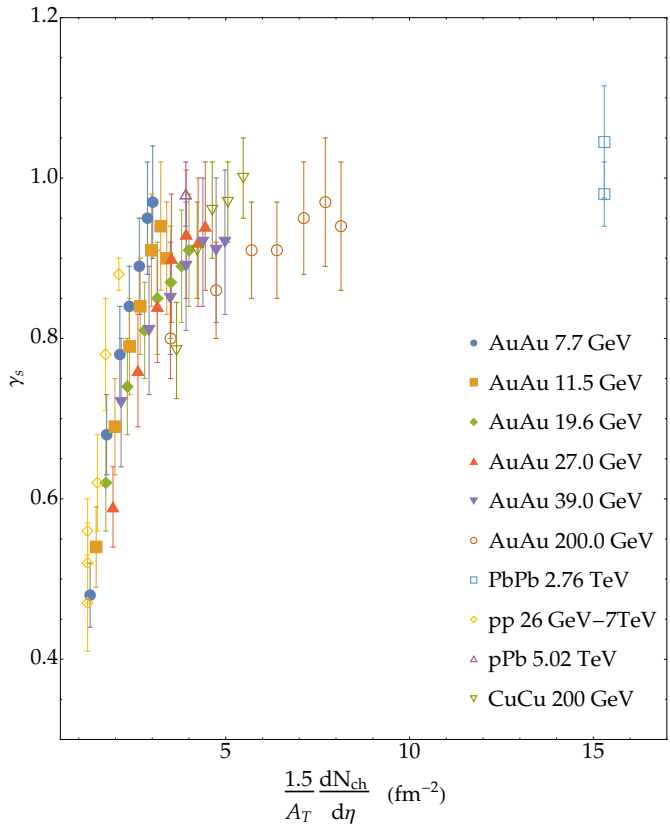}
	\caption{$\gamma_s$ universal behavior for data in~\cite{Becattini:1996gy,Takahashi:2008yt,Becattini:2012xb,Adamczyk:2017iwn}. Phobos parameterization~\cite{Alver:2010ck} for the relation between charge multiplicity, energy and the number of participants has been applied for RHIC data. For $pp$ and $pA$ the CGC parameterization of the transverse area as a function of multiplicity has been used~\cite{larry1,larry2,larry3}.} 
	\label{Fig:3}
\end{figure}

Fig.~\ref{Fig:3} from ref.~\cite{nuovo} clearly shows the universal pattern of $\gamma_s$, for different initial settings
$pp$, $pA$, $AA$ at various energies,
 when plotted versus
the initial entropy density $s_0$,  which in the one-dimensional hydrodynamic
formulation~\cite{Bj} is given by
\begin{equation}
\begin{split}
s_0 ~\!\tau_0 &\simeq \frac{1.5}{A_T}\; \frac{dN^x_{ch}}{dy} =
\frac{1.5}{A_T}\;\frac{N_{part}^x}{2} \left. \frac{dN^x_{ch}}{dy}\right|_{y=0}\;,
\label{star0}
\end{split}
\end{equation}
with $x = pp$, $pA$, $AA$.
Here $A_T$ is the transverse area, $(dN^x_{ch} / dy)_{y=0}$ denotes the number of produced charged secondaries,
normalized to half the number of participants $N_{part}^x$, in reaction 
$x$.  The transverse area $A_T$ is correlated with the multiplicity and it can
be evaluated in $AA$ and $pA$ by Glauber 
Monte Carlo simulation~\cite{Loizides:2017ack}. The high multiplicity events  in $pp$, are associated with fluctuation of the transverse area, which, in the Color Glass Condensate, are indeed fluctuations of the initial gluon field configurations~\cite{larry1,larry2,larry3}.
In proton-proton, proton-nucleus and heavy ion collisions at high energies, high multiplicities, $\gamma_s \rightarrow 1$ and it becomes a universal function of $s_0 ~\!\tau_0 $ as shown in fig.~\ref{Fig:3}.

Notice that the strangeness saturation, with $\gamma_s \gtrsim  0.95$, requires $s_0 \ge 6$.

\subsection{Elliptic flow and partecipant eccentricity}

The scaling behavior in $pp$ and $AA$ of the ratio between the elliptic flow, $v_2$, and the participant eccentricity, $\epsilon_{part}$, has been shown in ref.~\cite{nuovo}. Indeed the plot of $v_2/\epsilon_{part}$
versus $x=dN_{ch}/S$, where $S$ is the transverse area associated with
$\epsilon_{part}$, shows a universal trend starting from $x \ge 2.5$,
i.e. $s_0 \ge 3.8$. Moreover the difference between the geometrical transverse
area $A_T$  (i.e. the overlapping almond shape in $AA$ collisions)  and $S$ is
crucial to obtain the smooth interpolation among $pp$ and $AA$ data. The
definition of $S$ becomes  meaningless in $e^+e^-$, since it is related to the event by event fluctuations of the projectile/target constituents. Since $S < A_T$ one needs, in general, a value of $s_0$ larger than that one evaluated by geometrical criteria. In other words, a flow-like effect requires  large enough parton density in the transverse area.

\section{\label{sec:2}Transverse parton density in  $e^+e^-$ annihilation }

According to previous discussion, an universal behavior emerges if the parton density in the transverse plane is used as the relevant dynamical variable to define the initial setting of the collisions
and if it is large enough.

Let us now study  this quantity  in  $e^+e^-$ annihilation at different energies and multiplicities, starting from some phenomenological indications.

 \begin{figure}	
	\includegraphics[width=\columnwidth]{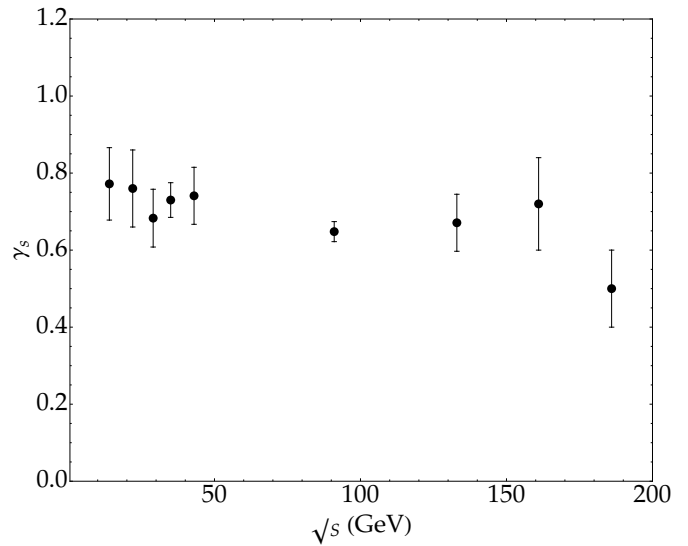}
	\caption{$\gamma_s$: energy dependence for $e^+e^-$ } 
	\label{Fig:4}
\end{figure}

The energy dependence of  $\gamma_s$~\cite{gammasnoi} in the range $14 - 186$ GeV for $e^+e^-$ is reported in fig.~\ref{Fig:4}. Within the error bars,  there is no signature of a enhancement of the strangeness production in the considered energy range: a clear difference with respect to  $pp$  and $AA$ collisions (see fig.~\ref{Fig:2}).

\begin{figure}	
	\includegraphics[width=\columnwidth]{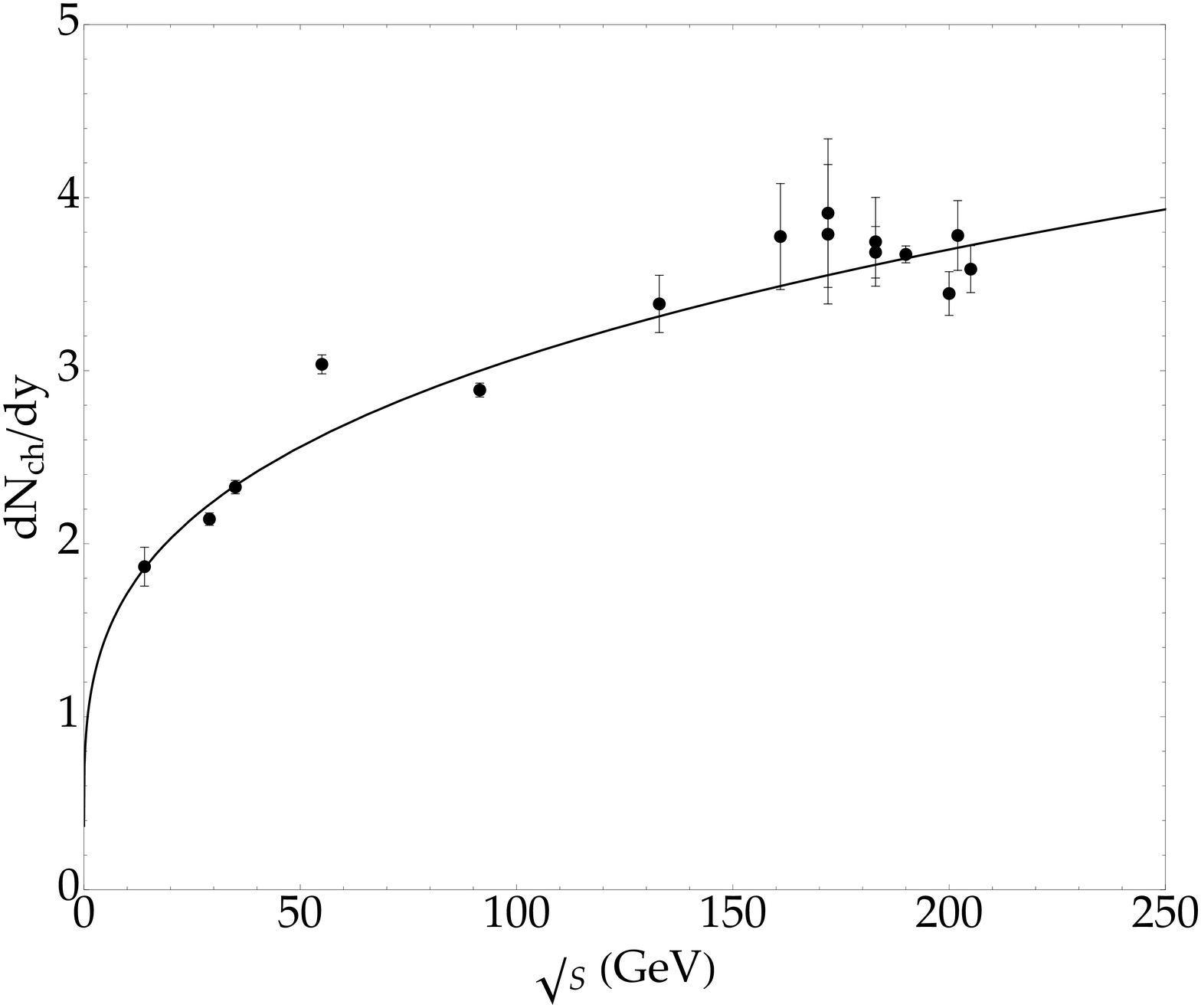}
	\caption{$dN_{ch}/dy$ in $e^+e^-$. Black line is the fit $\left(dN_{ch}/dy\right)^{e^+e^-}=0.3493+0.6837\; (\sqrt{s})^{0.3}$.} 
	\label{Fig:5}
\end{figure}

Therefore the question of a universality for $e^+e^-$ annihilation and of the
 saturation (if any) to $\gamma_s \rightarrow 1$ arises. 
To evaluate the effective parton density in the transverse plane for this
 particular non-hadronic setting, one has to know the multiplicity and the
 transverse area (which are not independent quantities). Indeed, the problem
 is a reliable evaluation of an effective transverse size for $e^+e^-$, since, in the energy range up to $\simeq 200$ GeV, the multiplicity is similar to the nucleus-nucleus one (normalized to half the number of participants $N_{part}$) and
$dN/dy$ at $y=0$, with respect to thrust axis, is plotted in fig.~\ref{Fig:5} versus the cms energy.

Let us recall, in  a simplified way, the steps of the hadronization cascade of a primary quark or antiquark produced in $e^+e^-$ annihilation. 

The color field flux tube (string), initially created along the direction of the separating $q$ and $\bar q$, produces a further pair $q_1$, $\bar q_1$ and then provides an increasing of their longitudinal momentum.
When $q_1$, $\bar q_1$  reaches the critical distances for  the string breaking  still another pair $q_2$, $\bar q_2$ is created 
and the state $q_2$, $\bar q_1$ is emitted as a hadron. 
The string multifragmentation produces the final multiplicity in fig.~\ref{Fig:5}.

\begin{figure}	
	\includegraphics[width=\columnwidth]{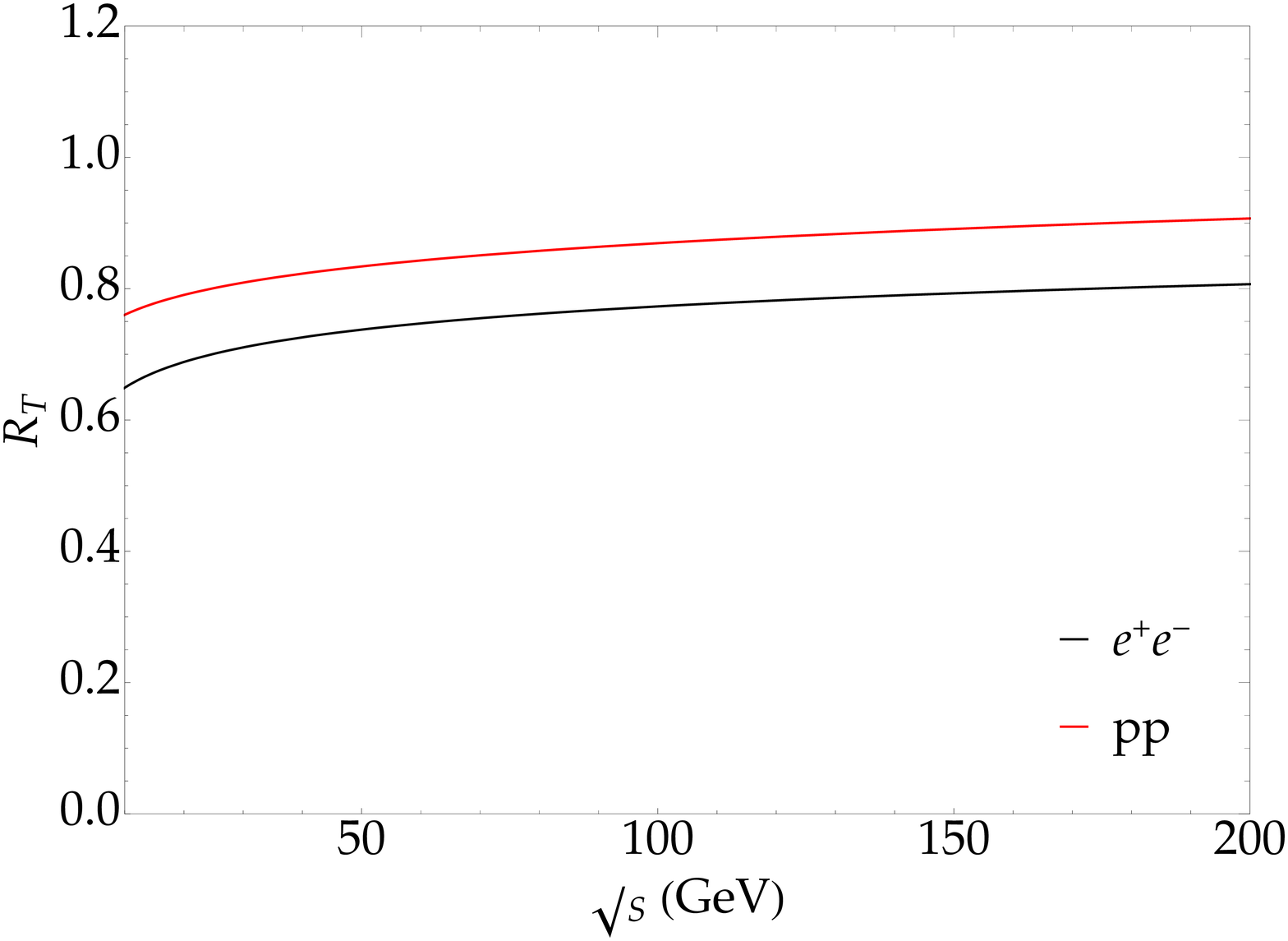}
	\caption{Transverse radius, $r_T$, in $e^+e^-$ (black) and $pp$ (red).} 
	\label{Fig:6}
\end{figure}

The lattice evaluation~\cite{Luscher:1980iy} of the transverse size, $R_T$, of a quark-antiquark string  at center mass energy $\sqrt{s}$ turns out to be~\cite{Luscher:1980iy, Castorina:2007eb}
\be
R_T^2 = \frac{2}{\pi \sigma} \sum_{k=0}^N \frac{1}{2k+1} 
\ee
where $\sigma$ is the string tension and $N = \sqrt{\pi s/ 2 \sigma}$. Moreover
\be
\sum_{k=0}^N \frac{1}{2k+1} = \frac{\gamma}{2} + ln(2) + \frac{1}{2}\left[\Psi(N+3/2)\right]
\ee
where $\gamma$ is the Euler-Mascheroni constant and $\Psi$ is the di-gamma function which, for large values of the argument, can be approximated as
\be
\Psi(x) \simeq ln(x) \;.
\ee
Finally the transverse size can evaluated by
\be
R_T^2 = \frac{2}{\pi \sigma} \left[\frac{\gamma}{2} + ln[2 \left(N+\frac{3}{2}\right)^{1/2}\right]
\ee
The result is plotted in fig.~\ref{Fig:6} and compared with the transverse size of a $pp$ collision, evaluated by the Color Glass parameterization, which takes into account the event-by-event fluctuations of the initial gluon configuration~\cite{larry1,larry2,larry3}, 
and by the phenomenological fit of the multiplicity
\be
\frac{dN}{dy}\Bigg|_{pp}= a_p + b_p \sqrt{s}^{0.22}
\ee
with $a_p=0.04123$ and $b_p = 0.797$~\cite{ada}.

The similarity between the transverse size in $e^+e^-$ and in $pp$ should not be surprising since it is well known that 
the multiplicity in $pp$ collisions is related with the multiplicity in $e^+e^-$ annihilation if one takes into account the leading particle effect, i.e.
the energy removed from the genuine hadronization cascade due to the leading particles~\cite{GrosseOetringhaus:2009kz,zic}.

\begin{figure}	
	\includegraphics[width=\columnwidth]{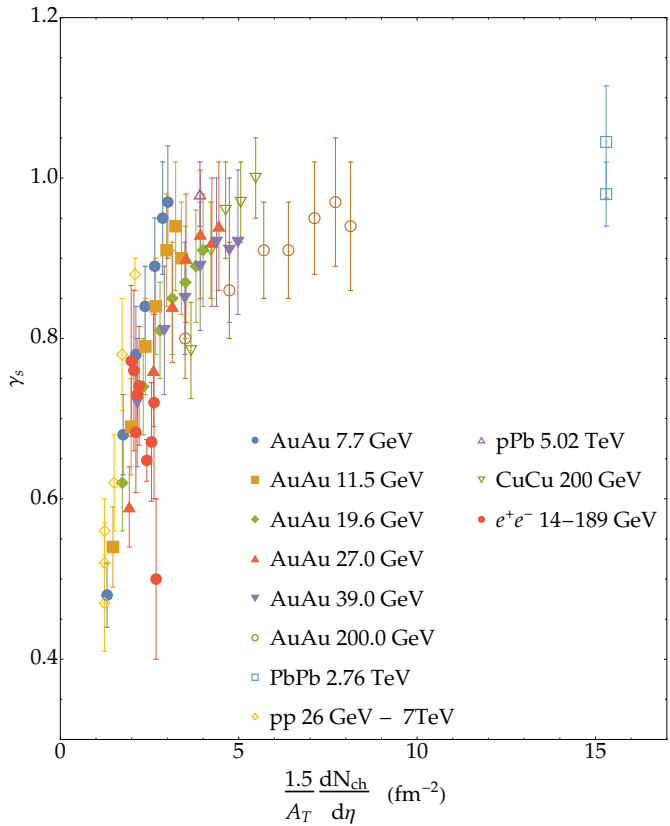}
	\caption{$\gamma_s$ versus $s_0$} 
	\label{Fig:7}
\end{figure}

The initial entropy density $s_0$ in $e^+e^-$ annihilation, can now be estimated by data in fig.~\ref{Fig:5}, fitted by $\left(dN_{ch}/dy\right)^{e^+e^-}=0.3493+0.6837\; (\sqrt{s})^{0.3}$, and by $R_T$ in fig.~\ref{Fig:6}. The energy range $14-186$ GeV correspond to rather narrow interval of $s_0$,
$ 2 \;\text{fm}^{-2} \lesssim s_0 \lesssim 3\;\text{fm}^{-2}$ and the $\gamma_s$ data in fig.~\ref{Fig:4} can be plotted on the universal curve as a function of $s_0$ for comparison with $pp$ and $AA$~(fig~\ref{Fig:7}).

\section{\label{sec:CC}Comments and Conclusions} 

The previous analysis clarifies that in $e^+ e^-$ annihilation at the LEP or lower energies there is no chance of observing the enhancement of the strangeness production, that is $ \gamma_s \gtrsim 0.95$,
because the parton density in the transverse plane is too small: $s_0$  turns out to be $\le 3$ but a value larger than $6$ is required.

The multiplicity one needs in  $e^+e^-$ annihilation to obtain the same value of $s_0$ determined for $pp$ and $Pb-Pb$ collisions is reported in table~\ref{TAB:1}, by extrapolating to very high energy
the fit of the multiplicity at lower energy in fig.~\ref{Fig:5}. The corresponding center-mass energy depends on the power of $\sqrt{s}$ in the fit and with previous formula for $dN_{ch}/dy$ one gets
$s_0=5.2 \pm 0.8$, corresponding to $dN/dy \simeq 10 \pm 2$, at $\sqrt{s} \simeq 7_{-4}^{+6}$ TeV.

\begin{table}
	\begin{tabular}{|c|c|c|c|c|}
		\hline
		$\frac{1.5}{A_T}\frac{dN_{ch}}{d\eta}$&
		$\left(\!\!\frac{dN_{ch}}{d\eta}\!\!\right)_{\!\!e^+\! e^-}\!\!$& $\left(\!\!\frac{dN_{ch}}{d\eta}\!\!\right)_{\!\!pp}\!\!$&$ \left(\!\!\frac{dN_{ch}}{d\eta}\!\!\right)_{\!\!PbPb}\!\!$&  \begin{tabular}{c}\footnotesize{$PbPb$}\\\footnotesize{Cent.}\end{tabular} \\
		\hline 
		$20.1\pm 0.8$& $58\pm 3$ & $100.\pm 4.$ &$ 1943.\pm 56.$ & \text{0-5$\%$} \\
		$17.5\pm 1.1$& $49\pm 4$ & $87.\pm 5.$ & $1587.\pm 47. $& \text{5-10$\%$} \\
		$15.4\pm 0.9$& $42\pm 3$ & $76.\pm 4.$ & $1180.\pm 31. $& \text{10-20$\%$} \\
		$12.2\pm 0.6$& $31\pm 2$ & $60.6\pm 3.1$ &$ 649.\pm 13.$ & \text{20-40$\%$} \\
		$8.3\pm 0.7$& $19\pm 2$ & $41.2\pm 3.4$ & $251.\pm 7. $& \text{40-60$\%$} \\
		$5.2\pm 0.8$& $10\pm 2$ & $26.\pm 4.$ & $70.6\pm 3.4 $& \text{60-80$\%$} \\
		$3.1\pm 1.1$& $5\pm 3$ & $12.4\pm 3.0$ & $17.5\pm 1.8$ & \text{80-90$\%$} \\
		\hline 
	\end{tabular}
	\caption{$dN_{ch}/d\eta$ in $PbPb$ at $5.02$~TeV, $pp$ and $e^+ e^-$ for different values of the variable in Eq.~\eqref{star0}.}
	\label{TAB:1}
\end{table}

Similarly, the value $s_0|_{e^+e^-}$ is too small for observing flow-like effect, although in this case a precise value is difficult to determine, due to the uncertainty in the estimate of the transverse area
associated with the eccentricity.

In conclusion, one can expect to see in $e^+e^-$ annihilation the asymptotic
behavior measured in high energy
$pp$ and $AA$ collisions only at very much higher energies than so far
available. There is a hierarchy in energy and multiplicity  requirements
to see the ``collective'' effects,
starting from low energy in $AA$ to larger energy and multiplicity in $pp$
collisions and still much larger energies in  $e^+e^-$ annihilation.

{\bf Acknowledgements}

P.C is partially supported by Charles University Research Center (UNCE/SCI/013).

\end{document}